\documentclass[aps,pra,twocolumn,10pt,floatfix]{revtex4-2}
\usepackage{graphicx,epsfig}
\usepackage{amsmath}
\usepackage{amsfonts}
\usepackage{fancyhdr}
\usepackage{listings}
\usepackage{float}
\usepackage{placeins}
\usepackage{hyperref}
	\hypersetup{
    colorlinks=true,
    linkcolor=blue,
    filecolor=blue,      
    urlcolor=blue,
    }

\usepackage{lipsum}

\begin{document}

\pagestyle{fancy}
\lhead{\bf Using games to tame complexity}
\rhead{Aleix Nicolás Olivé}

\title{Using Gamified Experiments to Tame Complexity: the case of the Schelling Model of Segregation}
\author{Aleix Nicolás Olivé}

\affiliation{Facultat de F\'{\i}sica, Universitat de Barcelona, Mart\'{\i} i Franqu\`{e}s 1, 08028 Barcelona, Spain.}
\author{Luce Prignano}
\affiliation{Departament de Física de la Matèria Condensada, Facultat de F\'{\i}sica, Universitat de Barcelona, Mart\'{\i} i Franqu\`{e}s 1, 08028 Barcelona, Spain.\\ Heuristica Lab.}

\author{Dimitri Marinelli}
\affiliation{Departament de Física de la Matèria Condensada, Facultat de F\'{\i}sica, Universitat de Barcelona, Mart\'{\i} i Franqu\`{e}s 1, 08028 Barcelona, Spain.\\ Universitat de Barcelona Institute of Complex Systems - UBICS}

\author{Emanuele Cozzo}
\email{emanuele.cozzo@ub.edu}
\affiliation{Departament de Física de la Matèria Condensada, Facultat de F\'{\i}sica, Universitat de Barcelona, Mart\'{\i} i Franqu\`{e}s 1, 08028 Barcelona, Spain.\\ Universitat de Barcelona Institute of Complex Systems - UBICS\\ CNSC Internet Interdisciplinary Institute, Universitat Oberta de Catalunya}

\date{\today}

\begin{abstract}
{\bf Abstract:}
This study employs gamified experiments to investigate and refine the Schelling Model of Segregation, a framework that demonstrates how individual preferences can lead to systemic segregation. Using a movement selection algorithm derived from a board game adaptation of the classical Schelling Model, the research examines player strategies aimed at minimizing segregation and maximizing happiness within a controlled environment. Rooted in greedy optimization, the model balances these objectives through a tunable parameter. Empirical data from gameplay is analyzed using Approximate Bayesian Computation, providing insights into player strategies and their alignment with systemic outcomes. The findings highlight the potential of gamification as a tool for engaging with complex social phenomena, enhancing agent-based models, and fostering participatory approaches in the study of emergent behaviors. This dual-layered framework incorporates collective decision-making into micro-macro models, addressing critiques of oversimplification and expanding their utility in educational and policy contexts.
\end{abstract}

\maketitle


\section{Introduction}
Micro-macro models \cite{raub2014introduction} are intended to explain social macro-level phenomena as a result of the behavior of the individual actors at the micro-level and their interactions. This research program has formal and methodological roots in mathematical sociology \cite{coleman1964introduction}, but set the foundations for a wide interdisciplinary field of research \cite{pnas2002}. Thanks to increasing computational capabilities, micro-macro models took on the appearance of agent-based models (ABM)\cite{helbing2012agent}, where individual actors' behavior could be specified both by means of equations or, more frequently, by decision rules. These models are capable of capturing the complexities of social systems, making them a powerful tool for exploring issues such as segregation, cooperation, and conflict \cite{Helbing2012}. Thus analytical analysis and simulations are used to characterize the emergent collective behaviours. This methodology resulted of particular appealing for statistical physicists \cite{castellano2009statistical}, for the wide applicability of statistical physics framework to connect the micro-level interactions to the macro-level emergent phenomena.

Several criticisms have been raised regarding this approach. Venturini et al. \cite{venturini2015filling} highlight a structural limitation of agent-based models (ABMs), where agents at the micro-level are typically portrayed as unable to comprehend or influence macro-level phenomena. Similarly, Jensen \cite{jensen2019politics} argues that this limitation leads physicists’ simple social models to adopt an external perspective, assuming that control and change in social systems must originate externally. This perspective aligns with top-down governance frameworks and risks reinforcing centralized control rather than empowering individuals within the system. Furthermore, he argue that, these models often oversimplify human behavior by treating individuals as "social atoms" with static, arbitrary characteristics, neglecting the dynamic and context-dependent nature of human interactions. 

Jensen also underlines the tendency in physics to "tame complexity" by simplifying real-world phenomena for laboratory experiments, making them more manageable and predictable. While this approach works well in physical sciences, social models attempting to gain a grasp on human behavior face the risk of "taming" humans in ways that may reduce their richness and unpredictability, as seen in systems like the social credit system in China. This can lead to better predictions but at the cost of ethical concerns and a reduction in the diversity of human behavior \cite{jensen2019politics}.

To address the kinds of criticisms often raised against simple sociophysical models, we developed a gamified experimental framework that combines the participatory engagement of role-playing games (RPGs) \cite{janssen2006empirically} with the abstraction and clarity of physical modeling. This framework allows participants to interact with simplified representations of social dynamics through structured games, introducing human agency and decision-making into the system. By enabling players to collectively solve problems and explore the relationship between individual actions and systemic outcomes, the approach addresses critiques of static agents and overly deterministic, top-down perspectives in traditional models.

Unlike the detailed and context-specific RPGs commonly used in social sciences \cite{janssen2006empirically}, which aim to replicate real-world scenarios, our framework retains the simplicity and generality characteristic of physical modeling by using board games. This ensures that the games are accessible and focused on capturing essential mechanisms of social dynamics while still producing meaningful emergent behaviors. Importantly, the data generated through these gamified experiments—such as participant strategies, decision-making processes, and observed outcomes—serve as a foundation for developing and refining new models. By systematically analyzing experimental results, we can create models that incorporate both the emergent dynamics observed in the games and the human factors often missing.

In contrast to attempts to "tame" social behavior, our framework embraces complexity by using interactive, participatory experimentation. Rather than simplifying or controlling human interactions, our approach allows participants to explore and engage with the dynamics of social systems. This enables the development of models that account for both individual agency, collective decision-making, and emergent systemic behavior, while reducing the arbitrariness left to the modeler. 

This participatory approach is further enriched through co-creation of the game with stakeholders, ensuring that the design process reflects diverse perspectives and practical needs. In that sense, our framework aligns with the participatory ethos of citizen science by actively involving stakeholders, such as teachers and other collaborators in our case, in the co-creation of the gamified experiments. This participatory design process ensures that the games are not only accessible and engaging but also tailored to address relevant educational and societal goals. Similar to co-creation approaches in citizen science, which emphasize collective decision-making and collaboration at all stages of the research process \cite{Senabre2018}, our framework uses participatory design to integrate stakeholder input into the development of the game. However, while participatory citizen science often focuses more on open-ended and context-specific processes \cite{Vohland2021}, our framework remains grounded in physics, simplifying systems to focus on their core mechanisms while maintaining the balance between simplicity and complexity.

To illustrate this framework, we present a specific example based on the Schelling model of urban segregation. In this board game, agents represent two ethnic groups, and the city is depicted as a grid. Each agent has a tolerance threshold: if the fraction of agents of the other type in their neighborhood exceeds this threshold, they become unhappy and move to an empty site where they are content. While the model's micro-level dynamics are based on the classical Schelling model, the game introduces a unique feature: players must cooperate within groups to minimize segregation, which is the most likely outcome of the dynamics, while still respecting the model's rules.

Building on empirical observations from gameplay, we propose a new model composed of two layers. The first layer models the micro-level dynamics, following the classical Schelling model, where agents move only when unhappy, seeking a location where they will be content. The second layer models the collective decision-making process, which is informed by the global state of the system and oriented toward collective values. Drawing from direct observations of player behavior and feedback after games (not reported here), we model this process as a greedy optimization algorithm.

We argue that this dual-layer approach addresses the criticism of oversimplification by preserving the essence of the micro-macro model. Unlike other models that encode the orientation toward a common good directly into the micro rules (see, for example, \cite{article1}), our framework introduces an explicit collective decision-making layer. This preserves the methodological and epistemological spirit of micro-macro models, while grounding the collective control modeling in empirical data collected from real gameplay, rather than relying solely on the modeler's assumptions.\cite{Szczepanska2022gam}.

\section{The model}
\subsection{The classical Schelling model of spatial segregation}
\label{Schelling}
In this subsection we briefly recall the definition of the classical Schelling model \cite{article3} for completeness.

A city is modelled as a $N\times N$ grid, whose sites $\sigma_i$ can be occupied by agents of one of two types, say red $\sigma_i=1$, and blue $\sigma_i=-1$, or can be empty $\sigma_i=0$. The fraction of empty sites is $\rho_0$. Agents have a tolerance threshold $\tau$, and are happy if the fraction of neighbors of the opposite type is below or equal to the threshold. Neighborhoods $\mathcal{N}(i)$ are defined as the eight sites surrounding a given site $i$. Starting from uniformly distributed agents, at each time step, a random unhappy agent is selected and relocated to an empty site if there she would be happy. The dynamic stop when all agents are happy, or unhappy agents have no empty sites where they will be happy. These stopping configurations are called blocked or frozen configurations. It can be shown \cite{Gauvin2009} that a Lyapunov function for the dynamics exists and guide the system to local minima.

We also introduce the main observables. That is, global segregation $\mathcal{S}$, defined as the fraction of pairs of neighbors of different types:

\begin{equation}
    \mathcal{S}=\dfrac{\sum_i\sum_{j\in \mathcal{N}_{i}}\sigma_{i}\sigma_{j}\delta (\sigma_{j}-\sigma_{i})}{\sum_i\sum_{j\in \mathcal{N}_{i}}\sigma_{i}\sigma_{j}};
\label{S}
\end{equation}

the global happiness $\mathcal{H}$, defined as the fraction of happy agents:

\begin{equation}
    \mathcal{H}=\dfrac{\sum_{i}h_{i}}{N(N-\rho_0)},
\label{H}
\end{equation}
where $h_i=1$ if agent at syte $\sigma_i$ is happy and $h_i=0$ otherwise; and finally we define the total number of steps $\mathcal{T}$ needed to reach a frozen configuration. 
Evaluations of the segregation outcomes of the classic Schelling model in the probability distribution of the entire outcome space of happiness and segregation show that the path-dependent process generates sharper segregation than expected by random configurations, and that segregation occur almost surely when happiness is maximized \cite{Jin2022}. 
\subsection{Board game Adaptation}
\begin{figure*}[!htbp]
    \centering
    
    \includegraphics[width=0.43\textwidth]
    {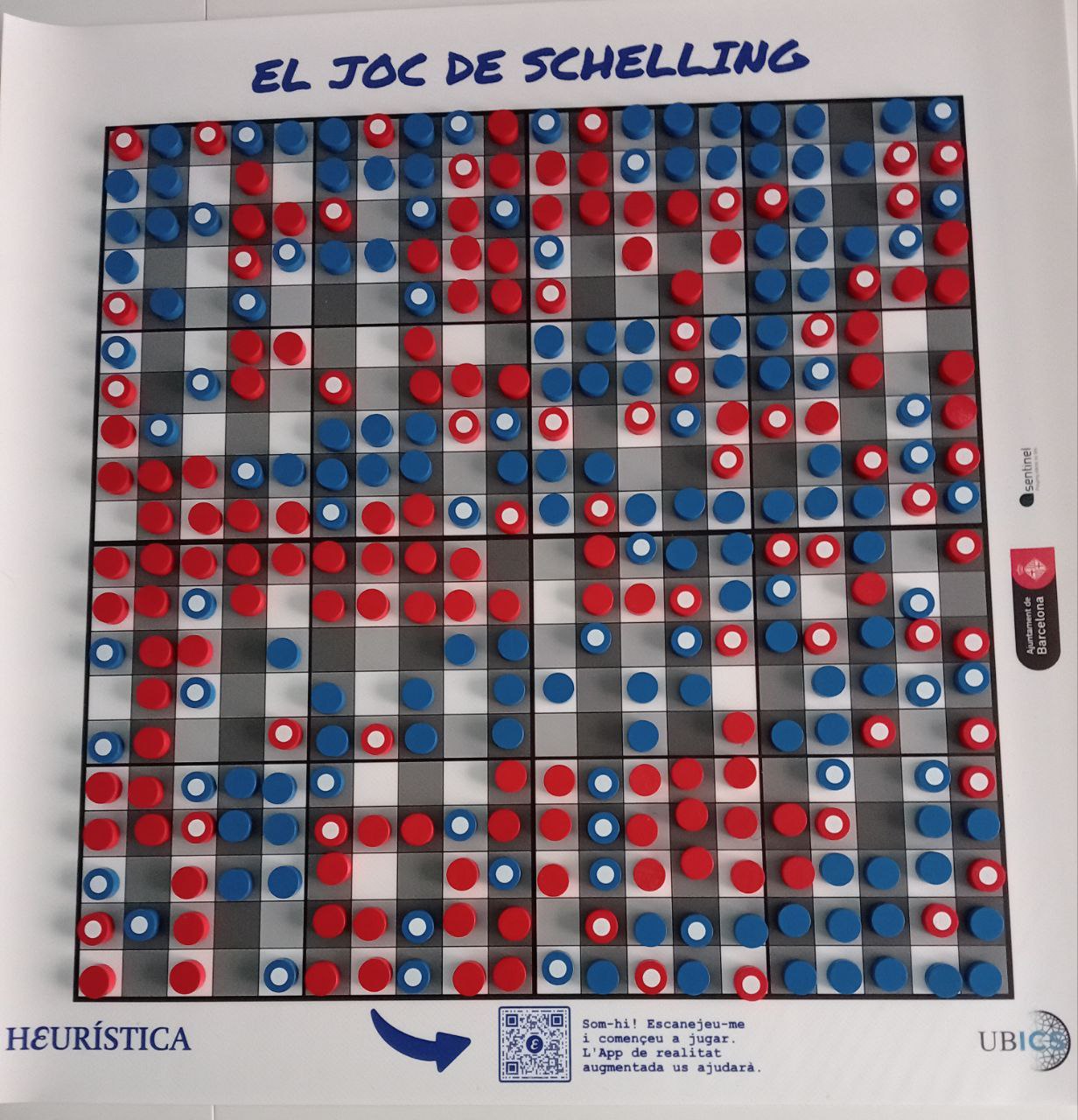} 
    \caption{Board game setup.}
    \label{fig:board}
\end{figure*}

The classic 2D Schelling model was adapted to a board game as part of an educational project on complex systems thinking by \textit{Associació Cultural Heurística}, with the game being co-created in collaboration with teachers and other stakeholders to ensure its relevance to educational and societal goals \cite{HeuristicaSite}. Given the educational purpose of the project, the parameters of the game version of the model have been selected to create a manageable and enjoyable experience while still exhibiting significant emergent behavior. The game consists of a $20\times20$ grid board with non-periodic boundary conditions. The fraction of empty sites is set to $\rho_{0}=0.2$, and there is an equal number of agents of types blue and red in the form of tokens (Fig \ref{fig:board}). Both types of agents have a tolerance threshold $\tau=0.5$, meaning they are considered happy if at least half of the agents in the occupied eight neighboring sites are of the same type.

This grid size $N=20$ and the fraction of empty sites $\rho_o$ are such that they allow for the characteristic dynamics of the Schelling model, achieving a frozen configurations where all agents are happy without the game becoming too long or cumbersome \cite{Gauvin2009}. The equal number of agents and the value of the threshold simplify the game. The selected threshold has not only been widely discussed in the literature, but also makes it easier for players to determine the happiness of the agents by assessing if a given site has more neighbors of one type or the other. It is also above the critical threshold \cite{Gauvin2009}, resulting in a final segregated state, which demonstrates how decisions at the micro-scale lead to unoptimal effects at the macro-scale.

\begin{figure*}[!htbp]
    \centering
    
    \includegraphics[width=0.43\textwidth]
    {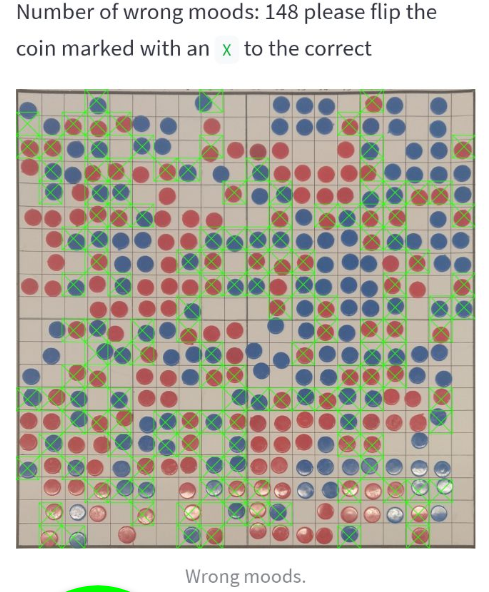} 
    \caption{Example of the helper app showing wrong happiness status.}
    \label{fig:mood}
\end{figure*}

A helper app was developed alongside the board game, allowing players to upload pictures of the board. These pictures are processed by a machine learning algorithm that classifies the agents at each site and determines their happiness, enabling players to easily update the happiness status of each agent (Fig. \ref{fig:mood}). 

\begin{figure*}[!htbp]
    \centering
    
    \includegraphics[width=0.43\textwidth]
    {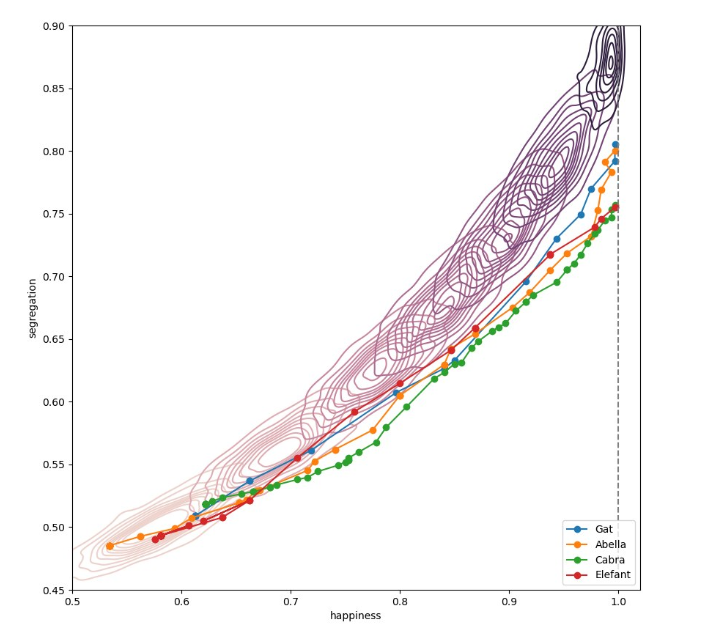} 
    \caption{Example of the helper app's visualization. The contour lines represent percentiles of the probability distribution of the classical Schelling model for a given happiness value. The dotted lines show real-time realizations of the gameplays.}
    \label{fig:viz}
\end{figure*}

The app also computes and visualizes the segregation value and happiness fraction, allowing players to monitor their progress (Fig \ref{fig:viz}).

The game works mostly like the classical Schelling model. The rules state that any unhappy agent can be moved, but only to a position where it would become happy. Players are grouped into teams of four. The game’s goal is to have all agents happy while minimizing segregation in the system. Teams compete to achieve this goal, with the team scoring the lowest segregation value winning the game.

In this way, the cooperation among the players of the same team simulates a collective decision-making process oriented toward achieving a shared common good, i.e., an integrated and happy final state. 

\subsection{Greedy optimization model}
 In our proposed model, an unhappy agent is chosen with uniform probability, and, while satisfying the condition that the agent is happy in the new location, that is selected to minimize the following free energy 

\begin{equation}
\mathcal{G}=\alpha \mathcal{S} -(1-\alpha)\mathcal{H}; \alpha\in[0,1]
\label{G}
\end{equation}

where $\mathcal{S}$ is the total segregation, and $\mathcal{H}$ is the global happiness, defined in subsection \ref{Schelling}. 

The parameter $\alpha$ determines the relative weight of $\mathcal{S}$ and $\mathcal{H}$ in the free energy function. With $\alpha=0$, minimizing the free energy is equivalent to maximizing the next step global happiness $\mathcal{H}$, while with $\alpha=1$, minimizing the free energy is equivalent to minimizing the next step segregation $\mathcal{S}$. Intermediate values result in an optimization problem that combines both measures. In this way, $\alpha$ can model different strategies followed by players. Note that the model with $\alpha=0$ is not equivalent to the classical Schelling model, since the new location is not sampled uniformly at the random among the allowed ones, but will be the one the minimize the creation of new unhappy agents, thus minimizing tipping events that are one of the main drivers in the emergence of segregation \cite{SchellingBook}. 

While players in the game could, in principle, adopt strategies that account for future states of the system and anticipate long-term consequences of their actions, the proposed model focuses on a greedy optimization approach. This means that agents make decisions based solely on the immediate minimization of the free energy function $\mathcal{G}$, without considering how their moves might influence the system over multiple steps. This simplification reflects the observed behavior in the gameplay and serves as a baseline for understanding strategic decision-making.

\section{Data and Methods}
\subsection{Data collection and processing}
The experimental data considered in this work consist of the configurations of four boards at different points during the game, including the final configuration with global happiness $\mathcal{H}=1$ and the initial random configuration. This information was obtained from pictures taken by the participants and uploaded to the helper app.\\

The games analyzed in this work were played by participants of the 2023 SMS-Social Modelling and Simulations workshop. Researchers, being familiar with complexity, emergent phenomena, and the Schelling model itself, were in a unique position to focus on the strategic elements of the game, which is the aspect we aim to model. Also, in an ideal scenario of scientifically informed policies, these are the profiles that policymakers would likely consult to understand the social dynamics at play and make informed decisions. Finally, among all the games played so far across various contexts and player profiles, these were the only ones that lasted long enough to reach the final frozen configuration.\\

For the four boards considered, labeled as A, B, C, and D, we have respectively 13, 28, 42, and 17 ordered configurations, including the initial condition and the final frozen configuration. From these configurations, we can directly determine the global happiness and segregation at each stage. 
Instead, the number of steps ($\Delta \mathcal{T}$) that have taken place between two consecutive configurations have to be estimated. We have that

\begin{equation}
\Delta \mathcal{T} \geq \dfrac{1}{2}\sum_{i}|\Delta\sigma_{i}|,
\label{DTbound}
\end{equation}
where $\Delta\sigma_i=\sigma_i(\tau+\Delta\mathcal{T})-\sigma_i(\tau)$. By definition, for a single step, only two values of the sum in \eqref{DTbound} will have changed. Specifically, the old position will change from $\sigma_{i}(\tau)=\pm 1$ to $\sigma_{i}(\tau + 1)=0$, and the new position will change from $\sigma_{i'}(\tau)=0$ to $\sigma_{i'}(\tau+1)=\pm 1$. The difference in the state value of a site $i$ at two subsequent time-steps is $\Delta\sigma_{i}=\sigma_{i}(t+1)-\sigma_{i}(t)$. The sum over all sites $i$ of the absolute difference of the state $\sigma_{i}$ will be 2 (hence the $\frac{1}{2}$ factor). Successive movements will work in the same way if certain conditions hold. If more than one step takes place, it is possible that a single agent moves more than once. While the right-hand side of \eqref{DTbound} only counts one movement, it serves as a lower bound for the number of steps. Fortunately, this circumstance is relatively uncommon because it is impossible for the same agent to move twice in a row; the first movement will leave it in a happy state, and changes around its neighborhood must occur before it can move again. Additionally, if an agent moves from a site and another agent of the same type moves into it, only one movement will be counted. This situation is unlikely since if an agent leaves a site, it typically remains unavailable for agents of the same type without a change in the neighborhood. Moreover, during most dynamics, there are usually a large number of unhappy agents and empty sites, so reoccupying the same site with the same type of agent between two pictures should be uncommon. Therefore, we approximate:

\begin{equation}
\Delta \mathcal{T} \sim \frac{1}{2}\sum_{i}|\Delta\sigma_{i}|
\end{equation}
It is important to consider that the experimental data, in addition to the aforementioned approximation, may have at least two possible sources of error. The primary source could be player error, where participants might inadvertently make invalid moves. The second potential source of error is misclassification by the machine learning algorithm of the helper app. However, prior tests (not reported) demonstrated good accuracy, and we verified that for each data point, the number of agents of each type remained constant. This means that for a misclassification to go undetected, another misclassification of opposite sign would have to occur simultaneously. While theoretically possible, it is highly unlikely, as no such misclassification has been detected for any board.

While using the number of steps as the independent variable may appear more intuitive, we have decided to use the global happiness value instead, as $\mathcal{H}=1$ defines the end of the dynamics. This will make all subsequent comparisons much easier as the simulated data for both the classic Schelling model and our greedy optimization model have a common initial and final value with the experimental values $\mathcal{H}(0)=\mathcal{H}_{0}$ and $\mathcal{H}(\mathcal{T})=1$. Another reason to avoid the number of steps as the independent variable is that its typical value is model dependent. 

\subsection{Approximate Bayesian Computation for parameter inference}
Our ultimate goal is to infer the strategy exploited by a team of players to keep segregation low while achieving maximal global happiness. As we model players' actions as a greedy optimization of the free energy \eqref{G}, we aim to infer the value of the parameter $\alpha$ given the data. In a Bayesian framework, this involves evaluating the posterior distribution of $\alpha$ given the data, expressed as $p(\alpha\mid \{\sigma_i(\tau)\})$, where ${\sigma_i(\tau)}$ represents the board configuration at each time step $\tau$. However, formulated this way, the problem is somewhat ill-posed because the state of the board is highly degenerate \cite{mantzaris2018examining}. This means that different micro configurations can correspond to the same macroscopic state as described by the observables of interest $\mathcal{S},\tau$, and $\mathcal{H}$. Thus, formally, the Bayes rule would read

\begin{equation}
    p(\alpha\mid \{(\mathcal{S}(\tau),\mathcal{H}(\tau),\tau)\})=\frac{p(\{(\mathcal{S}(\tau),\mathcal{H}(\tau),\tau)\}\mid \alpha )}{p(\{(\mathcal{S}(\tau),\mathcal{H}(\tau),\tau)\}))}.
    \label{bayesrule}
\end{equation}

The likelilhood $p(\{(\mathcal{S}(\tau),\mathcal{H}(\tau),\tau)\}\mid \alpha )$ in \eqref{bayesrule} results intractable, thus we estimate the posterior distribution using an Approximate Bayesian Computation (ABC) approach. ABC is a method used in Bayesian inference when the exact likelihood is difficult or impossible to compute\cite{article4}. To approximate the likelihood function a number of simulations with model parameter values sampled from the prior distribution are performed. Then, the model generated data is compared with the observed data by defining a distance metric. A threshold distance is set and each simulation is accepted if their distance to the observed data is smaller than the threshold. The counting of the values of $\alpha$ for the accepted simulations will give an empirical distribution that approximate the posterior.

To compare trajectories generated by the simulation and those coming from the games, we will exploit the fact that both starts at roughly the same value of $\mathcal{H}\sim 0.5$ and ends at the exactly same value of $\mathcal{H}=1$, thus spanning the same range of values in $\mathcal{H}$, while the range of values for $\mathcal{S}$ and $\tau$ could be very different. Thus, we align the empirical and simulated trajectories based on $\mathcal{H}$ and compare the values of $\mathcal{S}$ and $\tau$ at the corresponding points of $\mathcal{H}$, using the distance metrics \eqref{distanceS} and \eqref{distanceT},

\begin{align}
    \label{distanceS}
    \mathcal{D}_{\mathcal{\mathcal{S}}}^{i}=\dfrac{1}{N_{h}}\sum_{j=1}^{N_{h}}(\mathcal{S}_{i}(\tilde{h}_{j})-\mathcal{S}_{exp}(h_{j}))^{2},\\
    \label{distanceT}
    \mathcal{D}_{\mathcal{T}}^{i}=\dfrac{1}{N_{h}}\sum_{j=1}^{N_{h}}(\mathcal{T}_{i}(\tilde{h}_{j})-\mathcal{T}_{exp}(h_{j}))^{2};
\end{align}
where $N_h=\mid\{h_j\}\mid$ is the number of the different values of the global happiness $\mathcal{H}=h_j$ in the empirical trajectory, while $\tilde{h}_j$ is the value of happiness in the simulated trajectory that is closer to the empirical value $h_j$.
Using the two distances separately, on would obtain two posterior distribution $p(\alpha\mid \{\mathcal{S}(h)\})$ and $p(\alpha\mid \{\mathcal{T}(h)\})$. The joint posterior distribution can be defined in a number of ways, we have decided to use a new combined distance as a weighted sum of the two distances which have already been calculated:

\begin{equation}
    \mathcal{D}_{\mathcal{C}}^{i}=\omega\dfrac{\mathcal{D}_{\mathcal{S}}^{i}}{\mathcal{M}ax_i(\mathcal{D}^i_{\mathcal{S}})}+(1-\omega)\dfrac{\mathcal{D}_{\mathcal{T}}^{i}}{\mathcal{M}ax_i(\mathcal{D}^i_{\mathcal{T}})}
\end{equation}
where we normalize bot distances with the maximum distance attained between simulations and empirical trajectories. 
With this new combined distance the same process as with the non combined distances can be used to find the new joint posterior probability distribution.

We have considered values of alpha from 0 to 1 with an interval of 0.05. For each of them, we have performed a total of 10000 simulations starting from each board's initial configuration. 

\section{Results}

\subsection{Classic Schelling Model}
Starting at each board's initial configuration, we have performed 10000 simulations of the classic Schelling model where a random unhappy agent and a random empty site are selected with uniform probability. The movement will be accepted only if the agent is happy in the new position.\\
\subsubsection{Final segregation}
\begin{figure*}[!htbp]
    \centering
    \begin{tabular}{cc}
        \includegraphics[width=0.43\textwidth]{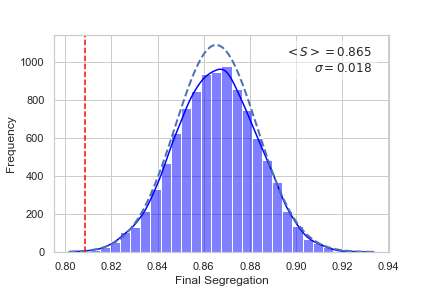} &
        \includegraphics[width=0.43\textwidth]{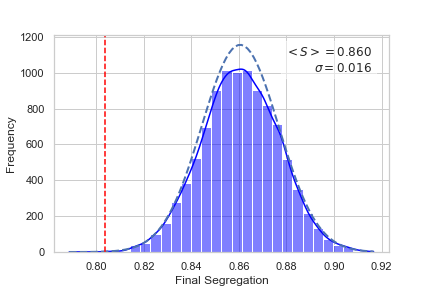} \\
        \includegraphics[width=0.43\textwidth]{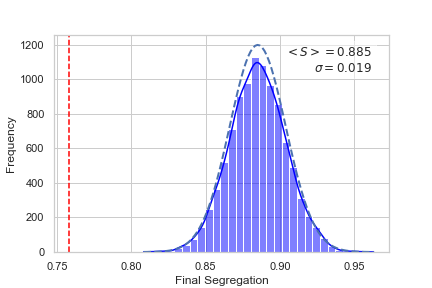} &
        \includegraphics[width=0.43\textwidth]{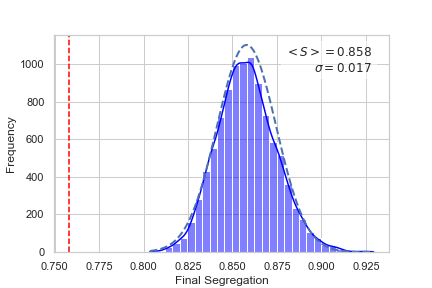} \\
    \end{tabular}
    \caption{Distribution of final segregation values for 10000 classic Schelling (histogram) with initial conditions corresponding to each board. Experimental value (Dashed red line), Gaussian fit (Dashed Blue line) and kernel density estimate of the probability distribution (solid blue line).}
    \label{fig:classic}
\end{figure*}
As seen in FIG. \ref{fig:classic}, the final values of segregation of the Classic Schelling simulations follow a Gaussian distribution with an expected value $\langle\mathcal{S}\rangle\in[0.86,0.89]$ depending on the initial conditions and a standard deviation $\sigma\in[0.016,0.019]$. All these expected values are relatively close and the standard deviation is also quite similar. For all initial conditions the experimental values are significantly lower than the expected value of the distribution.

\subsubsection{Final number of steps}
\begin{figure*}[!htbp]
    \centering
    \begin{tabular}{cc}
        \includegraphics[width=0.43\textwidth]{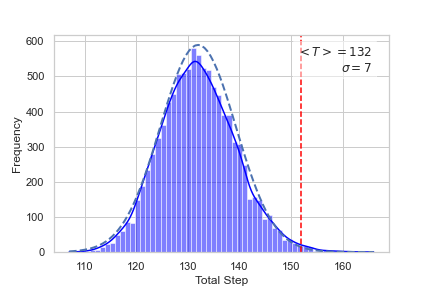} &
        \includegraphics[width=0.43\textwidth]{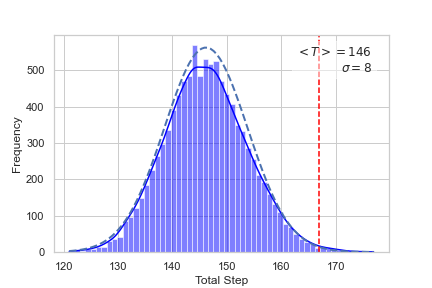} \\
        \includegraphics[width=0.43\textwidth]{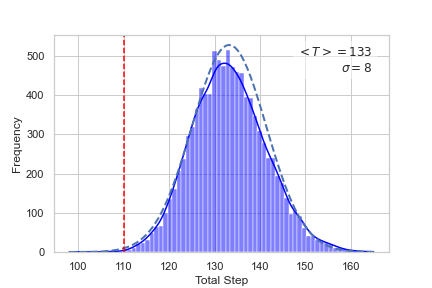} &
        \includegraphics[width=0.43\textwidth]{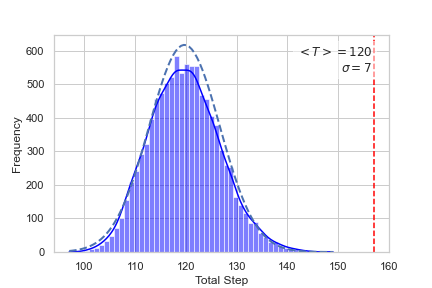} \\
    \end{tabular}
    \caption{Distribution of total number of steps for 10000 classic Schelling with initial conditions corresponding to each board. Experimental value (Dashed red line), Gaussian fit (Dashed Blue line) and kernel density estimate of the probability distribution(solid blue line).}
    \label{fig:final}
\end{figure*}

The final values of the number of steps also follow a normal distribution, as shown in FIG. \ref{fig:final}, with expected values $\mathcal{T}\in[120,146]$ and standard deviation with values between 7 and 8. There is a significantly higher relative difference between the expected values of the total steps compared to the final segregation values. All experimental number of steps are higher than the mean Schelling values except for board C.
\subsection{Behaviour of the greedy optimization model}
Here we explore the typical realization of the greedy optimization model for different values of the parameter $\alpha$.
\begin{figure*}[!htbp]
    \centering
    \begin{tabular}{cc}
        \includegraphics[width=0.43\textwidth]{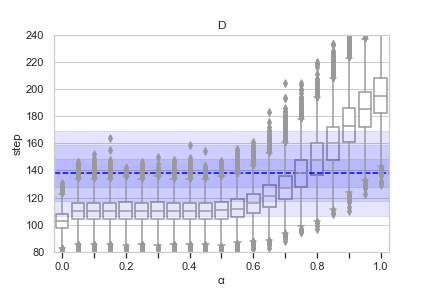} &
        \includegraphics[width=0.43\textwidth]{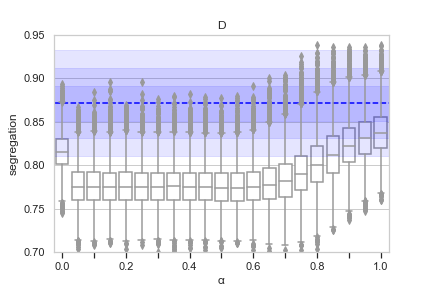} \\
    \end{tabular}
    \caption{Boxplot of total number of steps (left) and final segregation (right) as a function of the parameter $\alpha$ and mean of the classic Schelling model(dashed blue line) with 1,2, and 3 sigmas (shades of blue) for 10000 random initial conditions.}
    \label{fig:greedyrandom}
\end{figure*}
With respect to the final segregation, simulations show a clear dependence on the value $\alpha$ (see Fig \ref{fig:greedyrandom} right panel). The final segregation value is relatively high at $\alpha=0$. It displays a plateau of lower values from $\alpha=0.05$ to around $\alpha=0.6$, and from then on steadily increases.
The mean final segregation of the model is lower than that of classic Schelling model\\

\subsubsection{Final segregation}
\begin{figure*}[!htbp]
    \centering
    \begin{tabular}{ccc}
        \includegraphics[width=0.32\textwidth]{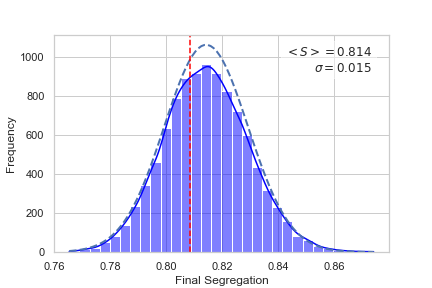} &
        \includegraphics[width=0.32\textwidth]{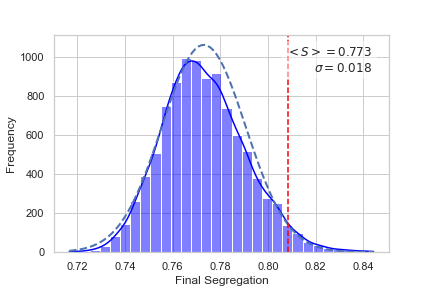} &
        \includegraphics[width=0.32\textwidth]{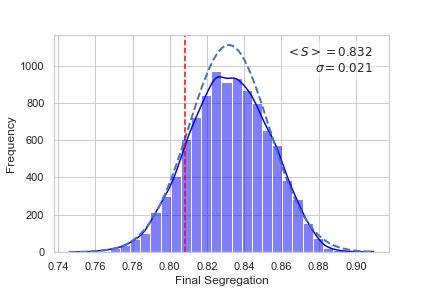} \\
        \includegraphics[width=0.32\textwidth]{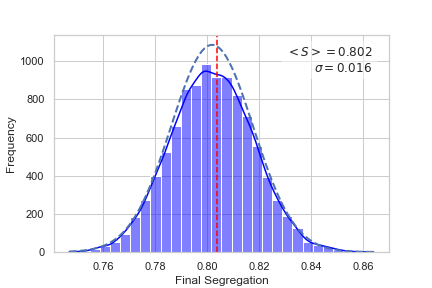} &
        \includegraphics[width=0.32\textwidth]{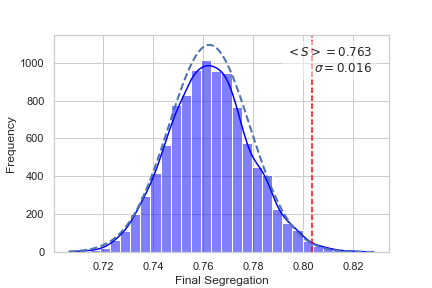} &
        \includegraphics[width=0.32\textwidth]{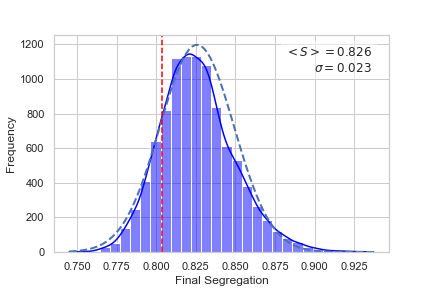} \\
        \includegraphics[width=0.32\textwidth]{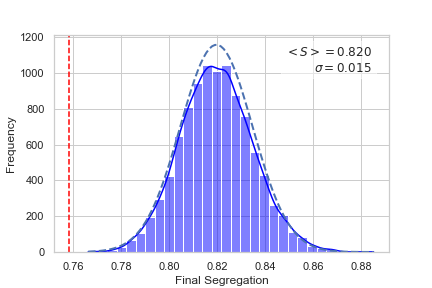} &
        \includegraphics[width=0.32\textwidth]{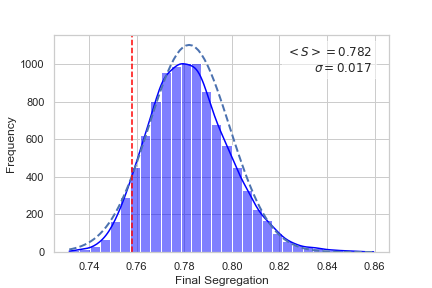} &
        \includegraphics[width=0.32\textwidth]{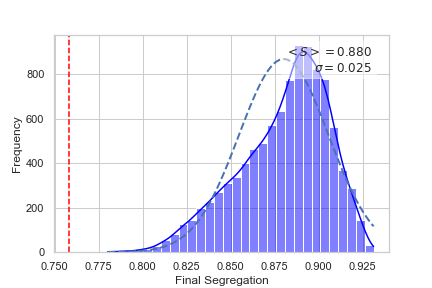} \\
        \includegraphics[width=0.32\textwidth]{Figures/Segregation_Distribution_alpha_0_Board_2.png} &
        \includegraphics[width=0.32\textwidth]{Figures/Segregation_Distribution_alpha_0.5_Board_2.png} &
        \includegraphics[width=0.32\textwidth]{Figures/Segregation_Distribution_alpha_1_Board_2.png} \\
    \end{tabular}
    \caption{Distribution of final segregation values for 10000 realizations (histogram) with initial conditions corresponding to each board. Experimental value (Dashed red line), Gaussian fit (Dashed Blue line) and kernel density estimate of the probability distribution (solid blue line).For the different boards (A,B,C,D from top to bottom) and different $\alpha$ values (0,0.5,1 from left to right)}
    \label{fig:segdistr}
\end{figure*}

\begin{figure*}[!htbp]
    \centering
    \begin{tabular}{cc}
        \includegraphics[width=0.43\textwidth]{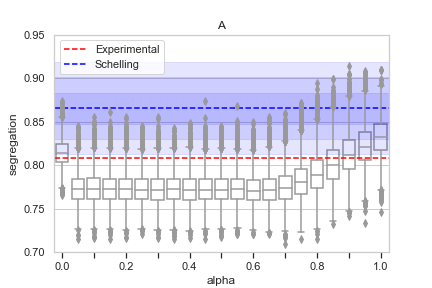} &
        \includegraphics[width=0.43\textwidth]{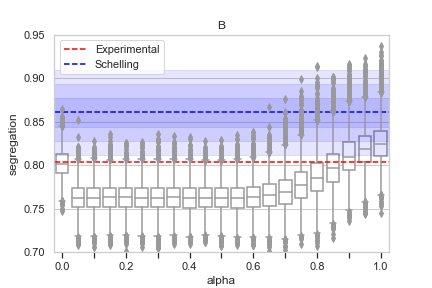} \\
        \includegraphics[width=0.43\textwidth]{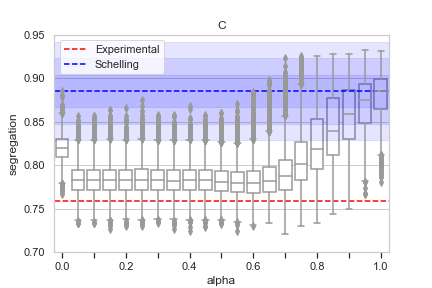} &
        \includegraphics[width=0.43\textwidth]{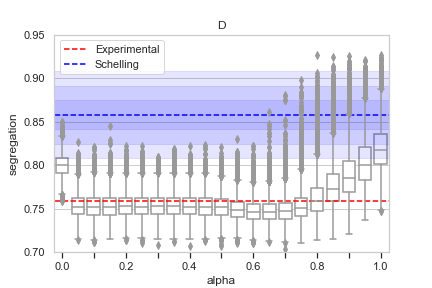} \\
    \end{tabular}
    \caption{Boxplot of final segregation as a function of parameter $\alpha$ as well as the experimental value (Dashed red line) and mean classic Schelling (Dashed blue line) value with 1, 2 and 3 sigmas (Shades of blue), for each board.}
    \label{fig:combined}
\end{figure*}

The distribution of final segregation values for our greedy optimization model, shown in FIG. \ref{fig:segdistr} for some values, does not always follow a normal distribution, specially for large values of $\alpha$ when they are clearly skewed when compared with the Gaussian fit.\\
We also observe in all boards that the segregation probability distribution gets wider as $\alpha$ increases. It is also noteworthy the fact that, for most boards and at all values of $\alpha$, there are more upper than lower outliers. This is only the opposite for high values of $\alpha$ of in board C.\\

The mean final segregation of the model is lower than that of Schelling for all boards and values of $\alpha$ except in the case of board C when, for $\alpha=1$, the values are similar. Even though they are lower there is significant overlap between both distributions, specially at higher values of $\alpha$.\\
For boards A and B the experimental value lies within the typical values of the model distribution for all values of $\alpha$ and within the interquartile range for $\alpha=\{0,0.85,0.9,0.95\}$.\\
Board C is the only one for which not only the median values but the lower quartile are above the experimental results for all values of $\alpha$. The experimental value lies within the typical values of the model distribution for all values of $\alpha\in [0.05,0.9]$ and specially close to the lower quartile for $\alpha \in [0.05,0.7]$ for higher values of $\alpha$ as well as $\alpha=0$ the experimental value is not within the typical distribution.\\
For Board D the experimental value lies within the typical values of the model distribution for all values of $\alpha$ except for $\alpha=0$ where it is lower. The experimental value lies on the higher quartile for $\alpha \in [0.05,0.5]$, above it for $\alpha \in [0.55,0.7]$ then just on the median for $\alpha=0.75$ and below the lower quartile for larger values of $\alpha$.

\subsubsection{Total number of steps}

\begin{figure*}[!htbp]
    \centering
    \begin{tabular}{ccc}
        \includegraphics[width=0.32\textwidth]{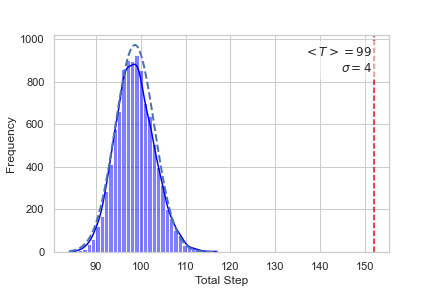} &
        \includegraphics[width=0.32\textwidth]{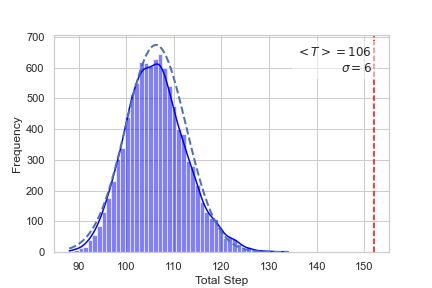} &
        \includegraphics[width=0.32\textwidth]{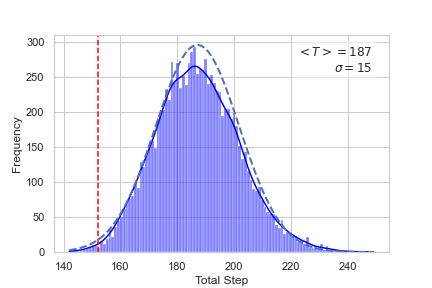} \\
        \includegraphics[width=0.32\textwidth]{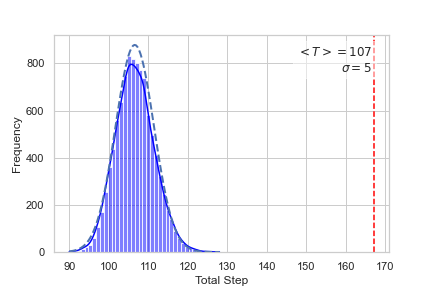} &
        \includegraphics[width=0.32\textwidth]{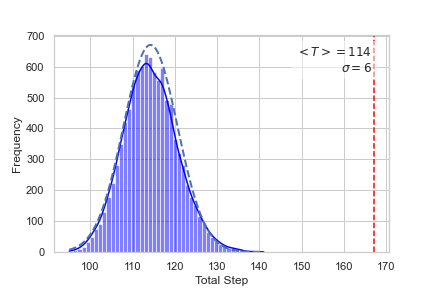} &
        \includegraphics[width=0.32\textwidth]{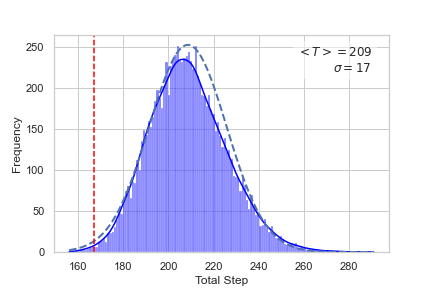} \\
        \includegraphics[width=0.32\textwidth]{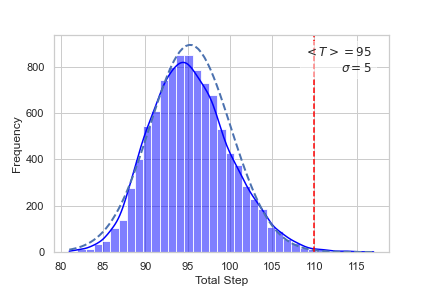} &
        \includegraphics[width=0.32\textwidth]{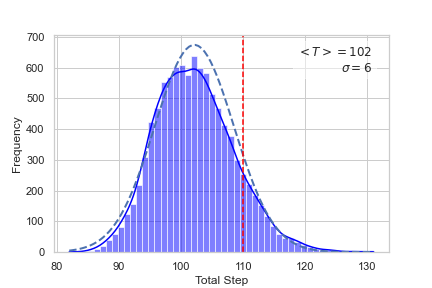} &
        \includegraphics[width=0.32\textwidth]{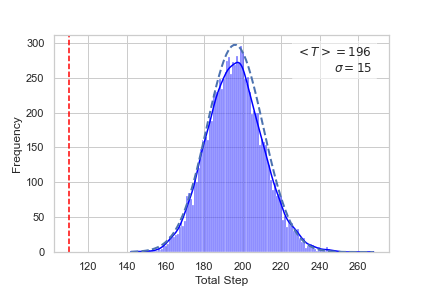} \\
        \includegraphics[width=0.32\textwidth]{Figures/Step_Distribution_alpha_0_Board_2.png} &
        \includegraphics[width=0.32\textwidth]{Figures/Step_Distribution_alpha_0.5_Board_2.png} &
        \includegraphics[width=0.32\textwidth]{Figures/Step_Distribution_alpha_1_Board_2.png} \\
    \end{tabular}
    \caption{Distribution of total number of steps for 10000 realizations (histogram) with initial conditions corresponding to each board. Experimental value (Dashed red line), Gaussian fit (Dashed Blue line) and kernel density estimate of the probability distribution (solid blue line). For the different boards (A,B,C,D from top to bottom) and different $\alpha$ values (0,0.5,1 from left to right)}
    \label{fig:stepdistr}
\end{figure*}

\begin{figure*}[!htbp]
    \centering
    \begin{tabular}{cc}
        \includegraphics[width=0.43\textwidth]{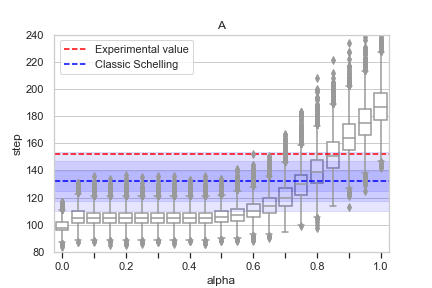} &
        \includegraphics[width=0.43\textwidth]{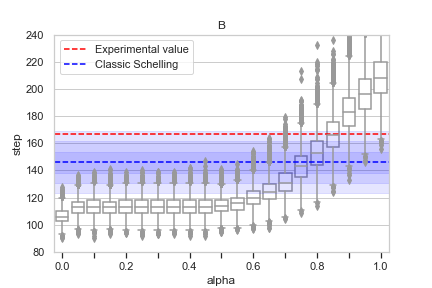} \\
        \includegraphics[width=0.43\textwidth]{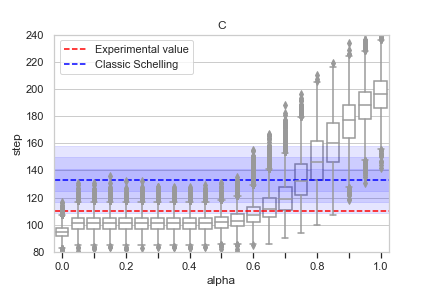} &
        \includegraphics[width=0.43\textwidth]{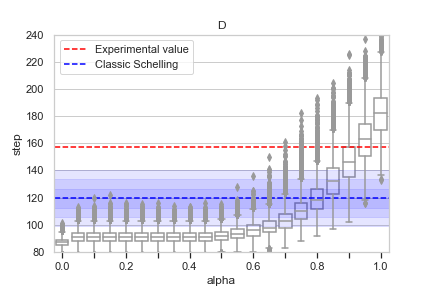} \\
    \end{tabular}
    \caption{Boxplot of total steps as a function of parameter $\alpha$ as well as the experimental value (Dashed red line) and mean classic Schelling (Dashed blue line) value with 1, 2 and 3 sigmas (Shades of blue), for each board.}
    \label{fig:combined}
\end{figure*}

The total number of steps distribution when using our algorithm in FIG 6. also shows some skew in comparison to the Gaussian fit.
Regarding the final number of steps, as seen in FIG. \ref{fig:stepdistr}  all boards show a similar behaviour. The parameter $\alpha=0$ corresponds to the lowest total number of steps, there is a slight increase in the total number of steps when $\alpha=0.05$ and stays stable until around $\alpha=0.5$ where the values start growing as $\alpha$ is increased. The total steps distribution also becomes wider at higher values of $\alpha$. This distribution also appears to have more and further away upper than lower outliers except, again, for high values of $\alpha$ in board C.\\

The experimental total number of steps is higher than the typical range of simulated values for board C when $\alpha\in [0,0.6]$ and lower for higher values of $\alpha$. It lies within the typical range for all values of $\alpha\in [0,0.65]$ and within the interquartile range for $\alpha=0.6$ and $\alpha=0.65$. The rest of the boards show a similar behaviour except for the fact that the experimental value is much higher relative to the simulations. The experimental value lays beyond the typical distribution for values with $\alpha<0.75$ or even $0.8$ in the case of board D. And are within the the interquartile range only for $\alpha=0.75,0.80$ for boards A and B. The only alpha for which it lays wihin the interquartile range of board C is $\alpha=0.8$. 

\subsubsection{Final segregation and total number of steps correlation}

\begin{figure*}[!htbp]
    \centering
    \begin{tabular}{cc}
        \includegraphics[width=0.43\textwidth]{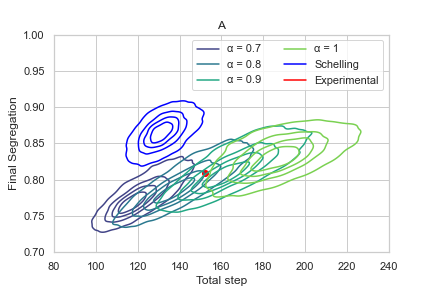} &
        \includegraphics[width=0.43\textwidth]{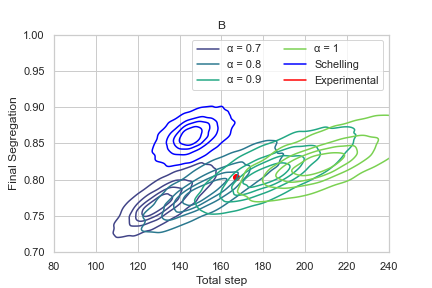} \\
        \includegraphics[width=0.43\textwidth]{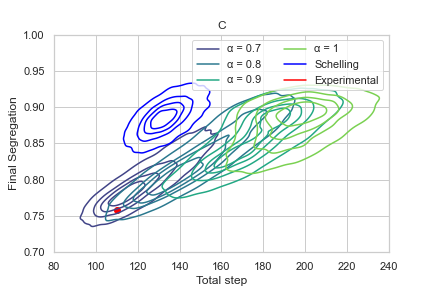} &
        \includegraphics[width=0.43\textwidth]{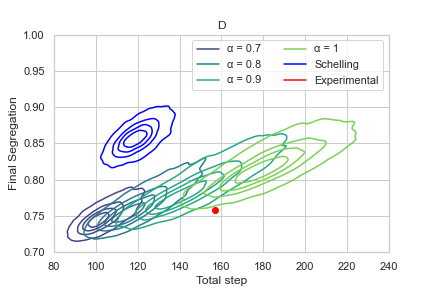} \\
    \end{tabular}
        \caption{Kernel Density Estimation of the Total Step-Final Segregation space for the classic Schelling model(Blue), and different values of the algorithm and the corresponding experimental value for each board.}
    \label{fig:dcombined}
\end{figure*}
In FIG. \ref{fig:dcombined} we can see how the kernel density estimation levels of different $\alpha$ values follow a clear tendency in the Total Step-Final Segregation space.  For all the boards the experimental value is never within the expected values of the Schelling model. For boards A and B the experimental value appears in a similar relative position to the $\alpha$ values represented while for c the experimental value is closer to lower values of $\alpha$. In Board D, however, the experimental value appears in the inferior right triangle instead of being within the expected values of any $\alpha$ value.\\
As we expected from the previous results, we can see that there is a positive correlation between the final segregation and the total number of steps as shown in FIG. 9. For all boards the correlation is smaller but stable for $\alpha\in [0,0.4]$ with correlation values ranging from 0.45 to 0.65, followed by a growth and peaking at around $\alpha=0.8$ and reaching correlation values around 0.8. This increase in correlation is probably due to the fact that at low values of $\alpha$ both, final segregation and final number of steps stay almost constant but at higher values of $\alpha$ both grow in an aproximately linear way.

\begin{figure}[!htbp]
    \centering
    \includegraphics[width=0.5\textwidth]{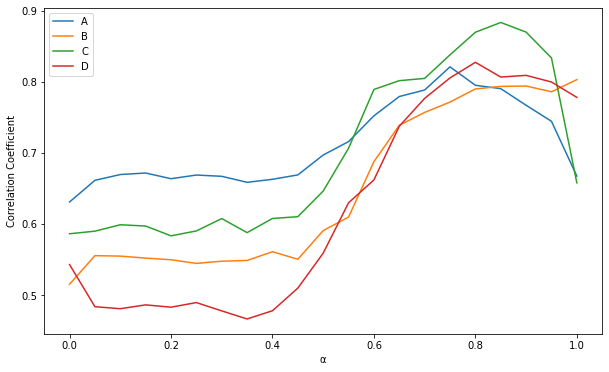}
    \caption{Correlation between Total Step and Final Segregation as function of $\alpha$.}
    \label{fig:sample}
\end{figure}

\subsubsection{Trajectory}
\begin{figure}[!htbp]
    \centering
    \resizebox{\columnwidth}{!}{
        \begin{tabular}{c}
            \vspace{-16pt}
            \includegraphics[width=0.7\columnwidth]{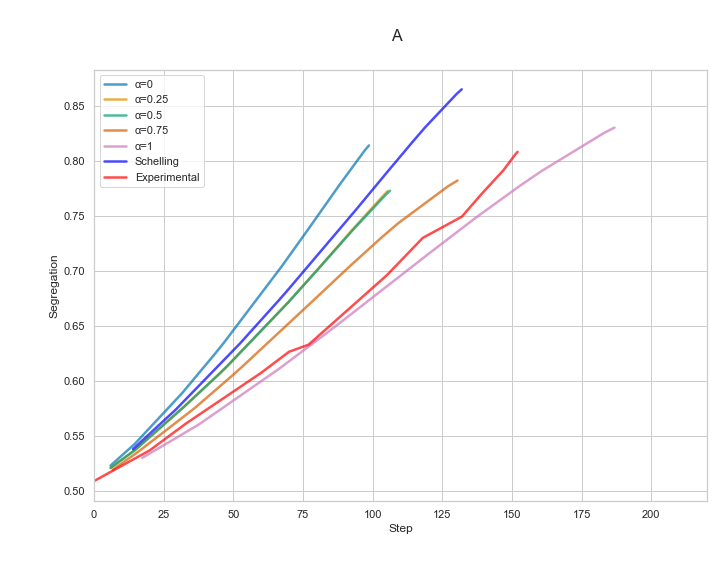} \\
            \vspace{-16pt}
            \includegraphics[width=0.7\columnwidth]{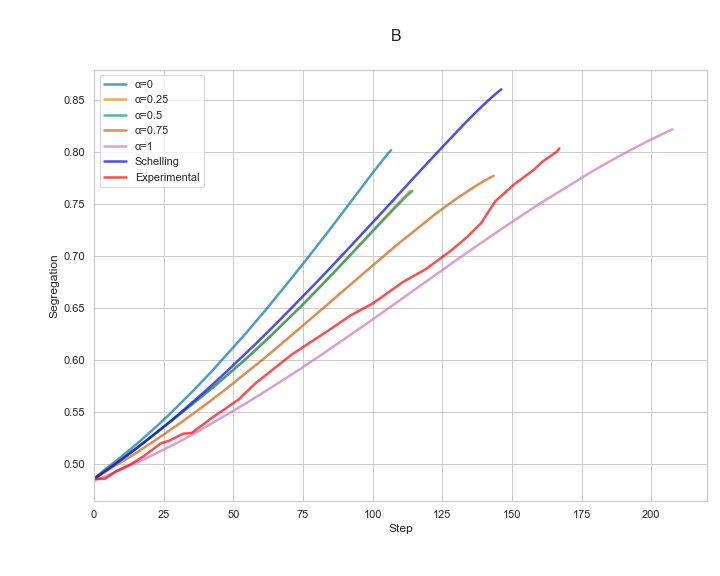} \\
            \vspace{-16pt}
            \includegraphics[width=0.7\columnwidth]{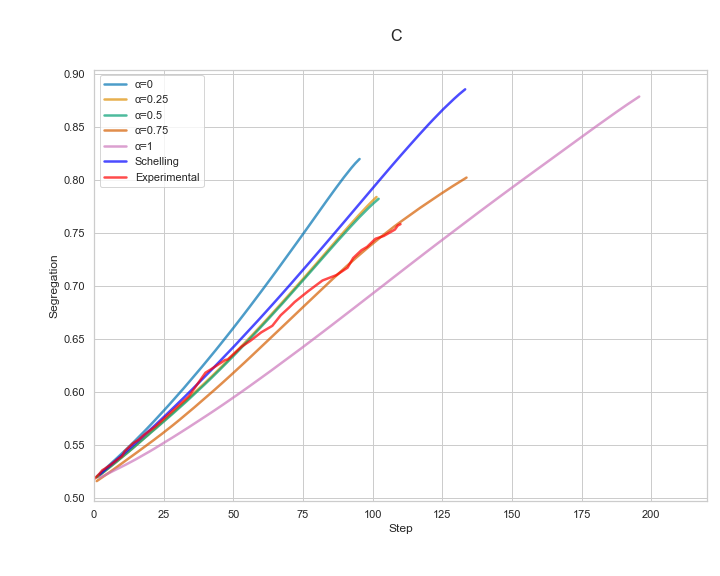} \\
            \vspace{-20pt}
            \includegraphics[width=0.7\columnwidth]{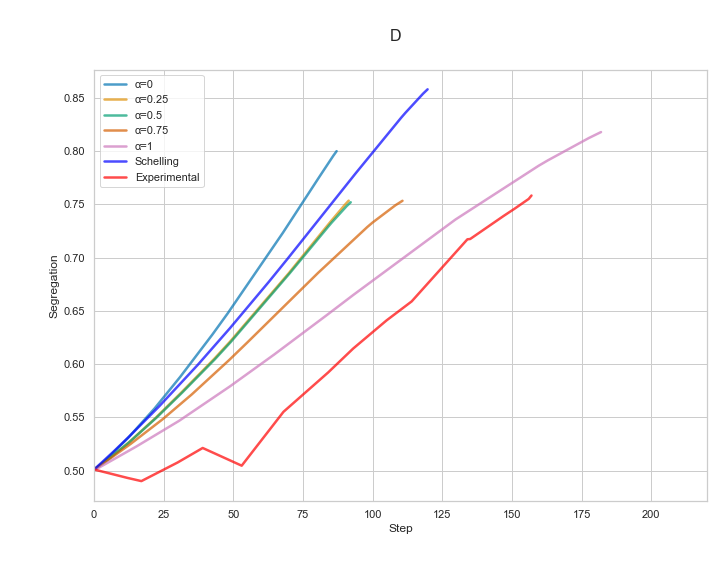} \\
        \end{tabular}
    }
    \caption{Average segregation as a function of the number of steps for different values of $\alpha$ and the classic Schelling model as well as the experimental values (red).}
    \label{fig:traj}
\end{figure}

\begin{figure}[!htbp]
    \centering
    \resizebox{\columnwidth}{!}{
        \begin{tabular}{c}
            \vspace{-15pt}
            \includegraphics[width=0.7\columnwidth]{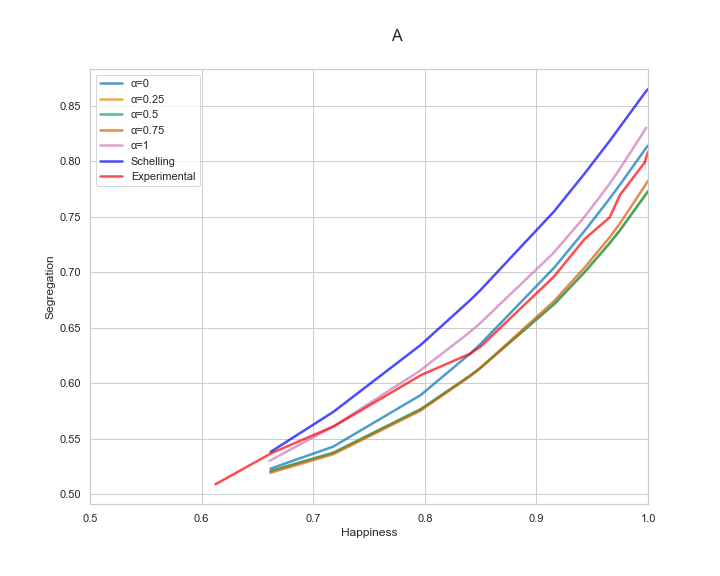} \\
            \vspace{-15pt}
            \includegraphics[width=0.7\columnwidth]{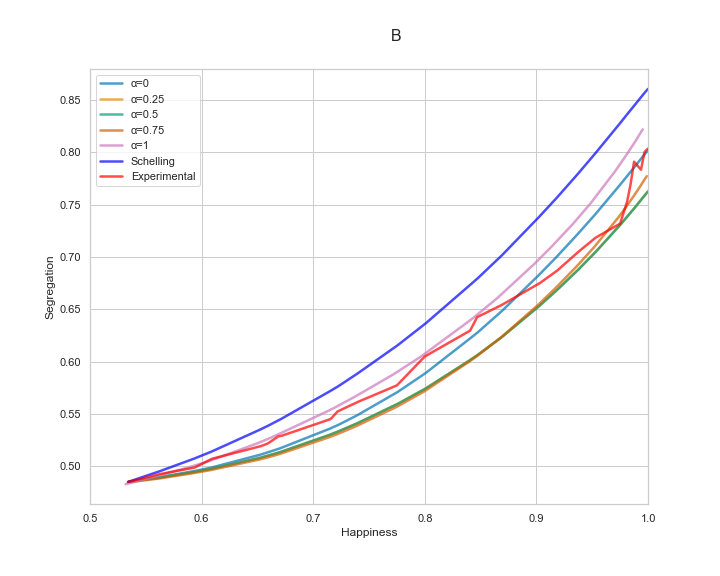} \\
            \vspace{-15pt}
            \includegraphics[width=0.7\columnwidth]{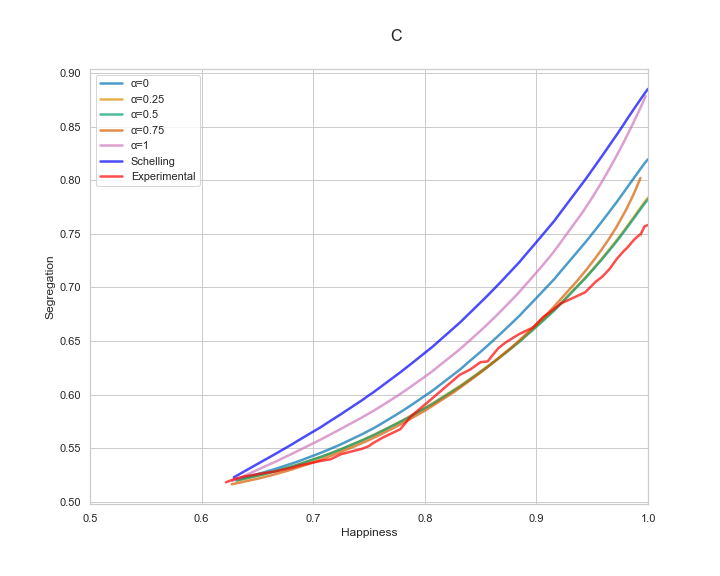} \\
            \vspace{-15pt}
            \includegraphics[width=0.7\columnwidth]{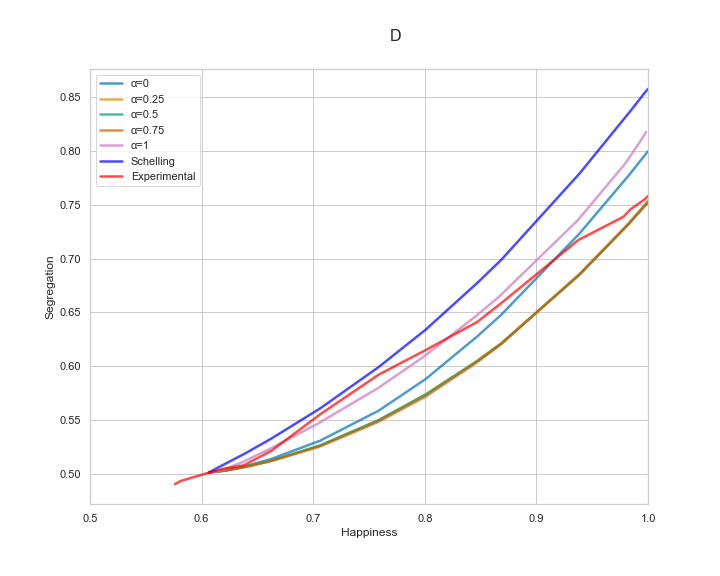} \\
        \end{tabular}
    }
    \caption{Average segregation as a function of the happiness for different values of $\alpha$ and the classic Schelling model as well as the experimental values (red).}
    \label{fig:traj2}
\end{figure}

When representing segregation as a function of the number of steps, as in FIG. \ref{fig:traj}, we can see that the growth is close to linear for the classic Schelling, all values of $\alpha$ and even experimental values. Therefore, we can say that the average segregation increase per step is constant. It's important to note that a higher slope does not necessarily translate into an increase in final segregation as they do not finish with the same number of steps. We can see that the growth rate diminishes as $\alpha$ grows with smaller values very close together. Experimental values can vary a little but usually stay between those of $\alpha=0$ and $\alpha=1$ except for the case D where the slope can take even negative values early on.\\
If we represent segregation as a function of happiness, like in FIG. \ref{fig:traj2}, we can see that the function is concave for the classic Schelling model, all values of $\alpha$ and the experimental values of boards A, B and C. This means that early on a large increase in happiness can be achieved with a relatively small increase in segregation while when the dynamics are close to the end the opposite is true. We can see how the evolution of the experimental values of board C runs close to low values of $\alpha$ but managed to achieve lower values towards high values of $\alpha$. Board D is also different than the other boards, as the slope remains mostly constant throughout the game. For this case the initial segregation values where relatively high at lower happiness value but dropped in comparison to the other series reaching relatively low segregation at the end.

\subsubsection{Posterior distribution}
We estimated the posterior probability distribution by following the Approximate Bayesian Inference methodology previously described.\\
We have defined the threshold to determine the posterior probabilities that only the $5\%$ smallest distance values are considered. This allows for a sufficiently large number of values for a statistical study without increasing excessively the noise. The robustness of this value has been confirmed by checking that other values close to this do not modify significantly the distribution.
The weight used to compute the combined distance has been set to $\omega=0.5$ which means that the same weight was put on both number of steps and segregation. We have also checked different weight values and found the results to be robust.

\begin{figure}[!htbp]
    \centering
    \includegraphics[width=0.5\textwidth]{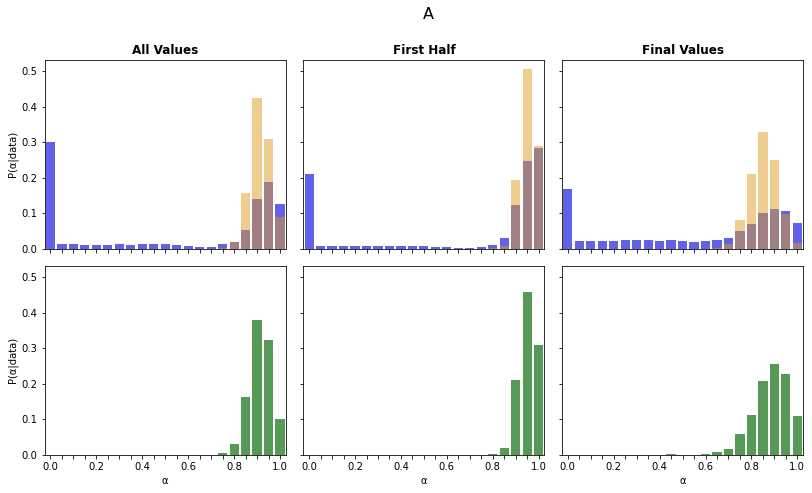}
    \caption{Board A. Posterior probability distribution obtained from segregation and steps (Top, Step in Orange and Segregation in Blue) and corresponding joint posterior probability distribution (Bottom in Green) considering: all the values (Left), the first half of the values (Center) and only the last value (Right).}
    \label{fig:sample1}
\end{figure}

\begin{figure}[!htbp]
    \centering
    \includegraphics[width=0.5\textwidth]{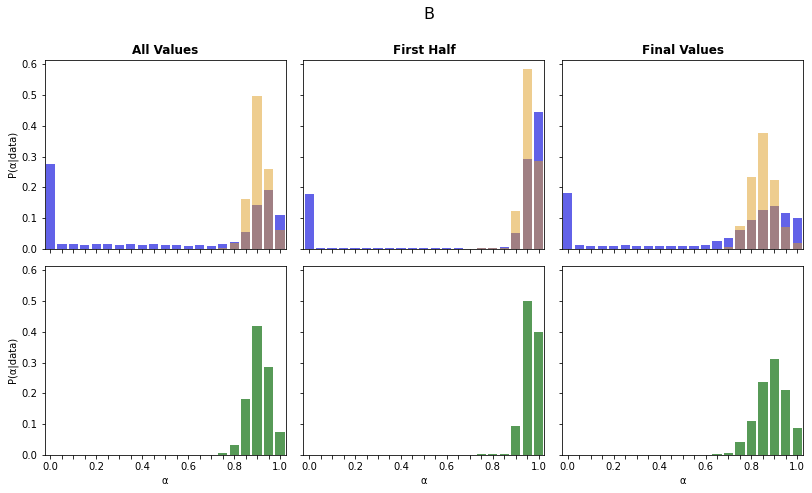}
    \caption{Board B. Posterior probability distribution obtained from segregation and steps (Top, Step in Orange and Segregation in Blue) and corresponding joint posterior probability distribution (Bottom in Green) considering: all the values (Left), the first half of the values (Center) and only the last value (Right).}
    \label{fig:sample2}
\end{figure}

\begin{figure}[!htbp]
    \centering
    \includegraphics[width=0.5\textwidth]{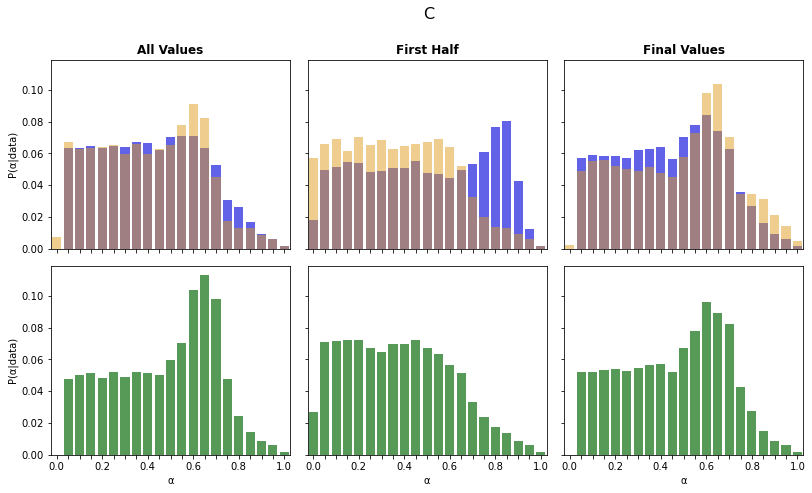}
    \caption{Board C. Posterior probability distribution obtained from segregation and steps (Top, Step in Orange and Segregation in Blue) and corresponding joint posterior probability distribution (Bottom in Green) considering: all the values (Left), the first half of the values (Center) and only the last value (Right).}
    \label{fig:sample3}
\end{figure}

\begin{figure}[!htbp]
    \centering
    \includegraphics[width=0.5\textwidth]{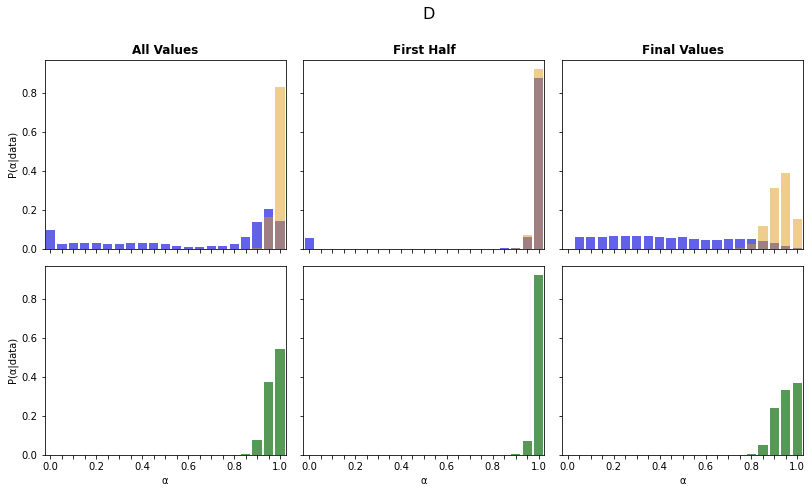}
    \caption{Board D. Posterior probability distribution obtained from segregation and steps (Top, Step in Orange and Segregation in Blue) and corresponding joint posterior probability distribution (Bottom in Green) considering: all the values (Left), the first half of the values (Center) and only the last value (Right).}
    \label{fig:sample4}
\end{figure}

The posterior distributions for boards A (FIG. \ref{fig:sample1}) and B (FIG. \ref{fig:sample2}) appear very similar. When using the segregation a high peak appears at $\alpha=0$ and a Gaussian like peak at high values of $\alpha$. The distribution at high values is more peaked when considering all values than when only considering the final ones, in the case where only the first half of the values have been considered is extremely peaked and the maximum is at $\alpha=1$.\\
If we consider now the total number of steps value posterior distribution we can see that the peak at $\alpha=0$ does not appear at all while a peak close to the one of segregation does appear. As in the segregation case, the peak is centered at higher values if we take into account only the first half of the values.\\
For both boards the posterior distribution obtained from the combined distance show a single peak, centered at 0.9 when considering all values or only the final value and at 0.95 when considering only the first half of the values.\\
The posterior distributions obtained from board C (FIG. \ref{fig:sample3}) are significantly different. The posterior from the combined distances for the first half are mostly uniform for a range $\alpha\in [0.05,0.6]$ and from there rapidly falls. However, considering all values, or the first half, a peak centered at $\alpha=0.65$ appears.\\
The posterior distributions obtained from all values and half of the values appear similar to those of boards A and B but with the peak at even higher values of alpha. When considering only the final values however, the posterior obtained from the segregation slowly diminishes with $\alpha$.
The posterior distributions for the first half of the values of board D(FIG. \ref{fig:sample4}) is very peaked and centered on a value of $\alpha=1.00$. However, when considering only the final values the posterior obtained from the segregation is wider and centered around $\alpha=0.95$ while the distribution obtained from the number of steps is much more uniform.\\
The posterior distributions using absolute displacement instead of mean squared displacement have also been computed and show comparable results.

\section{Discussion and Conclusions}
The introduced greedy optimization model demonstrated distinct differences compared to the classical Schelling model. By strategically minimizing a function that combines segregation and the opposite of mean happiness, the model consistently produced outcomes with lower typical segregation levels across all values of the parameter $\alpha$. This is likely due to its ability to reduce tipping events, which are a key driver of segregation in the classical model. These results underscore how incorporating strategic decision-making into agent behavior can significantly alter the emergent dynamics of the system.
The parameter $\alpha$ balances segregation $(\mathcal{S})$ and happiness $(\mathcal{H})$ within the greedy optimization model. By adjusting $\alpha$, the model shifts the relative importance of minimizing segregation versus maximizing happiness. When 
$\alpha=0$, the optimization prioritizes happiness, reflecting a strategy that avoids creating new unhappy agents. Conversely, when $\alpha=1$, the model focuses entirely on reducing segregation, directly targeting the intended systemic outcomes. Intermediate values of the parameter create a balance between these objectives, allowing the model to explore a spectrum of strategies. This flexibility demonstrates how $\alpha$ can represent diverse decision-making approaches, significantly shaping the emergent dynamics and final configurations.
The Approximate Bayesian Computation (ABC) analysis provided valuable insights into the strategies adopted by players during the game. The joint posterior distribution of the parameter $\alpha$ revealed a strong preference for values above $0.6$, indicating that players predominantly prioritized minimizing segregation over maximizing happiness. However, the model shows that for lower values of $\alpha$, the final segregation levels could have been reduced further, suggesting that the strategies employed by players were suboptimal in terms of achieving minimal segregation.
An exception to this pattern is observed in the results from board C, where the final segregation achieved by the players was lower than what the model predicts for any value of $\alpha$. The flatness of the posterior distribution for this board suggests a rejection of the model for the strategy followed by that team. This outcome highlights a case where the players employed strategies that deviated significantly from those captured by the model, demonstrating the potential for more sophisticated or unanticipated decision-making approaches that the proposed model does not fully encompass.

The findings demonstrate how the integration of collective decision-making into the model addresses criticisms of top-down assumptions often associated with traditional sociophysical models. By incorporating a second layer that explicitly models the collective optimization process, the framework moves away from imposing systemic outcomes directly through predefined micro-rules. Instead, it allows for emergent systemic behavior to arise from the interplay between individual decisions and group strategies. This approach preserves the methodological spirit of micro-macro models while introducing a more participatory element that aligns with observed behaviors in gameplay.

In contrast to purely top-down frameworks, where the modeler prescribes the orientation toward a common good, this dual-layer structure ensures that collective decision-making emerges from the dynamics of the system itself. The observed preference for strategies prioritizing segregation reduction (higher $\alpha$) highlights how players collectively align with broader systemic goals, even within the constraints of a simplified model. This design choice addresses the critique that sociophysical models often neglect the agency of individuals and groups in influencing macro-level outcomes, emphasizing instead how collective values and decision-making processes can be empirically integrated into the modeling framework.
The combination of gamified experiments with agent-based models (ABMs) represents a significant methodological innovation, particularly in its ability to leverage empirical data from gameplay for refining theoretical models. By collecting detailed data on player decisions and strategies during the game, the framework provides a rich empirical basis for calibrating ABMs. This approach allows for the systematic integration of observed behaviors into the model, reducing the reliance on arbitrary assumptions and enhancing the model's alignment with real-world decision-making processes. For example, the posterior distributions obtained through ABC analysis highlight the strategic tendencies of players, offering insights into collective behaviors that can inform refinements to the model's structure and parameters. Additionally, this methodology also allows for detecting behaviors not captured by the model, as demonstrated by the outcomes of board C.
This study has several limitations that should be acknowledged. First, the greedy optimization approach, while providing useful insights into decision-making strategies, inherently simplifies the problem of balancing segregation and happiness. The model assumes that agents will always select the most immediate optimization strategy, which may not reflect more nuanced, long-term planning that could occur in real-world scenarios. This simplification might limit the model’s applicability in cases where long-term consequences are significant or when players consider factors beyond immediate outcomes, as seen in more complex decision-making processes. Moreover, there are constraints within the experimental setup that could affect the generalizability of the results. The participant profiles, which largely included researchers and a few students. This limited sample size and the specific demographic of participants may introduce biases in the strategies observed and influence the outcomes of the gameplay. 
There are also challenges in generalizing the findings to different social contexts or scales. While the framework demonstrates utility in understanding segregation dynamics at a local level, scaling the model to larger or more diverse social systems may require adjustments to the parameters or game design. These factors must be considered when interpreting the results and planning for future applications of the framework.

\FloatBarrier

\end{document}